# Imaging the quantum melting of Wigner crystal with electronic quasicrystal order

Zhongjie Wang[1,2], Meng Zhao[1,2], Lu Liu[1,2,3], Chunzheng Wang[1,2], Fang Yang[4,5], Hua Wu[1,2,3,6]* & Chunlei Gao[1,2,4,5,6,7#]

*1 State Key Laboratory of Surface Physics and Department of Physics, Fudan University, Shanghai 200438, China*

*2 Shanghai Qi Zhi Institute, Shanghai 200232, China*

*3 Laboratory for Computational Physical Sciences (MOE), Fudan University, Shanghai 200438, China*

*4 Institute for Nanoelectronic Devices and Quantum Computing, Fudan University, Songhu Rd. 2005, Shanghai 200438, China*

*5 Zhangjiang Fudan International Innovation Center, Fudan University, Shanghai 201210, China*

*6 Collaborative Innovation Center of Advanced Microstructures, Nanjing University, Nanjing 210093, China*

*7 Shanghai Research Center for Quantum Sciences, Shanghai 201315, China*

*These authors contributed equally: Zhongjie Wang, Meng Zhao, Lu Liu*
*Corresponding authors: *wuh@fudan.edu.cn, #clgao@fudan.edu.cn*

**Wigner crystal, as the most fundamental exemplification where the many-body interaction forges the electrons into a solid[1,2], experiences an intriguing quantum melting[3-6] where diverse intermediate phases are predicted to emerge near the quantum critical point[7-13]. Indications of exotic Wigner orders like bubble phase[14], liquid-solid phase[15], and anisotropic Wigner phase[16] have been established by optical or transport measurements. However, the direct visualization of lattice-scale melting order, which is of paramount importance to unequivocally uncover the melting nature, remains challenging and lacking. Noting that Wigner crystals have been achieved in the fractionally filled moiré superlattice recently[17-20], here, via scanning tunneling microscope, we image the quantum melting of Wigner solid realized by further varying the moiré superstructure in monolayer $YbCl_3$/graphene heterostructure. The Wigner solid is constructed on the two-dimensional ensemble of interfacial electron-hole pairs derived from charge transfer. The interplay between certain moiré potential and ionic potential leads to the quantum melting of Wigner solid evidenced by the emergence of electron-liquid characteristic, verifying the theoretical predictions[8,9,11-13,21]. Particularly, akin to the classical quasicrystal made of atoms, a dodecagonal quasicrystal made of electrons, i.e., the Wigner quasicrytal, is visualized at the quantum melting point. In stark contrast to the incompressible Wigner solid, the Wigner quasicrytal hosts considerable liquefied nature unraveled by the interference ripples caused by scattering. By virtue of the two-dimensional charge transfer interface composed of monolayer heavy electron material and graphene, our discovery not only enriches the exploration and understanding of quantum solid-liquid melting, but also paves the way to directly probe the quantum critical order of correlated many-body system.**

In a dilute many-electron system, strong Coulomb correlation can localize and freeze electrons into periodic arrays forming a Wigner solid, in order to minimize the electrostatic energy which exceeds the motional quantum fluctuation[2]. The common strategy to quench the electron kinetic energy is to apply a magnetic field and enforce electrons filled into the Landau cyclotron orbits, whereby the close affinity between Wigner crystal and integer/fractional quantum hall states has been uncovered[22]. In the past several years, a novel strategy regarding moiré physics arises as filling electrons inside the moiré trapping potential efficiently suppresses the kinetic energy and enhances the comparative Coulomb correlation. Hence a series of fascinating states have been discovered: magic-angle superconductivity[23], correlated insulating state[24], quantum anomalous Hall state[25], and Wigner crystal settled in moiré lattices[17,19].

In electron solid immersed in a uniform neutralizing background, increasing the electron density leads to comparable kinetic energy competing with the electrostatic energy, resulting in the quantum melting of Wigner crystal[3-6]. Enigmatic strongly correlated orders are theoretically and experimentally reported to appear at the quantum critical melting point, manifested as the solid-liquid hybrid, nematic/hexatic Fermi liquids, and the hybrid of striped and bubbled orders[7-12,21,26]. Evidently, putting the mutually repulsive electrons in a composite electrostatic potential consisting of both the ionic and moiré potential, in replacement of the uniform electrostatic background, would greatly intertwine different physical interactions and totally reshuffle the game, leading to a Coulomb frustration regime[21,27]. An enlightening example illustrating the final arrangement of electrons in the composite scenario is the direct visualization of the Wigner crystal where the electrons are settled according to the dominating moiré landscape[17]. The moiré superstructure imposes a deep and intricate impact on filled tenuous electrons or interlayer excitons due to the similar spatial scale of the moiré unit cell and interelectron distance. As a result, varying the moiré parameter could also give rise to corresponding delocalized orders[28-31] besides the Wigner solid phase. What's more, embedding and constraining the electron solids in discrete lattice sites of another periodic spatial structure incommensurately is predicted to lead to the emergence of electronic quasicrystal[32,33]. Therefore, with the multifaceted involvement of electron correlation, ionic lattice, and moiré periodicity, the waning and waxing of the three parameters would introduce new complexity into the quantum phase diagram, appealing for the exploration of novel states subjected to competitive/collaborative mechanisms.

In this work, we fabricate monolayer-YbCl₃/graphene heterostructures via molecular beam epitaxy (MBE). The Wigner solid states, which stem from the two-dimensional electron-hole pairs of equilibrium induced by interfacial charge transfer, are visualized by scanning tunneling microscope (STM) at 5 K. The electron-hole pairs featuring Rydberg-like wavefunctions are identified by their spatial distribution normal to the surface. Various Wigner orders, including the melting phase of Wigner quasicrystal which is the electron counterpart of classic quasicrystal made of atoms[34], are observed in different stacking domains, indicating the intriguing interplay between Coulomb interaction, ionic lattice, and moiré periodicity.

## Charge transfer states inside YbCl₃ band gap

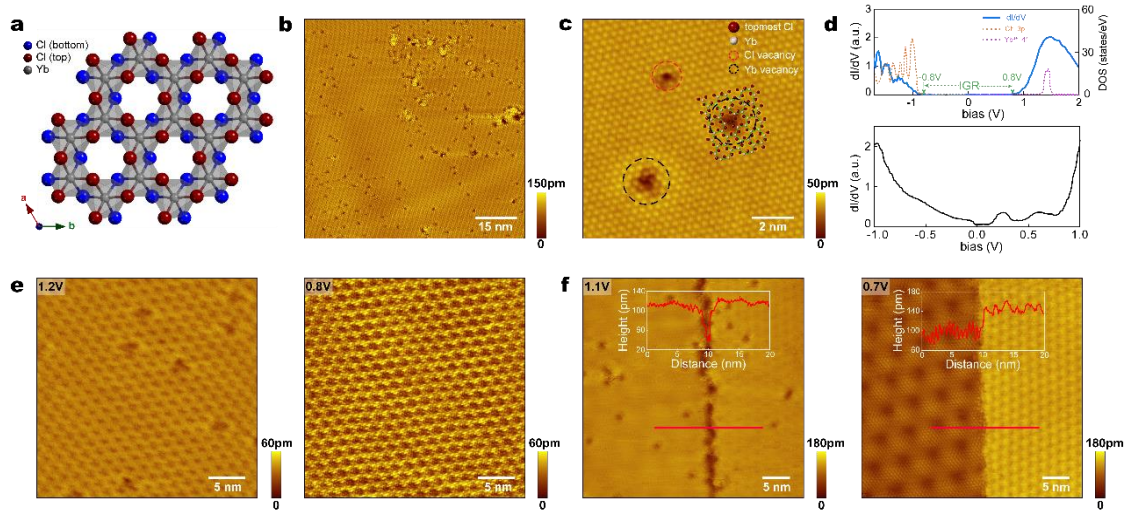

**Fig. 1. The topographic and electronic characteristics of YbCl₃/graphene heterostructure. a**, schematic illustration of crystal structure for monolayer YbCl₃. **b**, STM image for monolayer YbCl₃ of full coverage taken at U=1.2 V and I=100 pA. **c**, close-up equipped with atomic resolution, upon which the atomic structure and the type of defects can be identified and are indicated by the superimposed Yb- and topmost Cl-lattice plot. The red or black dashed circle points out the topmost Cl-defect or Yb-defect, respectively. (setup condition: U=2.0 V, I=100 pA) **d**, typical tunneling spectra measured on YbCl₃ monolayer. The top panel shows the dI/dV spectrum (blue curve) obtained at a large tip-sample distance (setup condition: U=2.1V, I=100 pA) together

with the computed density of states of YbCl$_3$ monolayer (the Yb 4f- and Cl 3p-bands are denoted as violet and orange dotted curves respectively). The bottom panel shows the tunneling spectrum obtained at a small tip-sample distance (setup condition: U=1.0 V, I=100 pA). **e**, two nanoscale STM images taken at the same position but at different bias voltages, illustrating a superlattice which exhibits a $\sqrt{7} \times \sqrt{7}$ superstructure with respect to the YbCl$_3$ unit cell. The defects imaged at 1.2 V (left panel) disappear when lowering the bias to 0.8 V (right panel). (Setup condition: U=1.2 V and I=500 pA for left panel, U=0.8 V and I=500 pA for right panel.) **f**, STM images at the intersection of two domains which present different superlattices. The two images are taken at the same position but at different bias. The insets show the corresponding height profiles along the horizontal red line (Setup condition: U=1.1 V and I=10 pA for left panel, U=0.7 V and I=10 pA for right panel). Compared to the left panel where two domains show the same height and clearly resolved defects, a noticeable height difference develops and the defects disappear in the right panel.

The rare-earth halide YbCl$_3$ is a representative strongly correlated material of 4f electrons[35]. Fig. 1a illustrates its atomic structure where the honeycomb layer of Yb-ions is sandwiched between two layers of triangularly arranged Cl-ions. Single layer YbCl$_3$ is epitaxially grown on the substrate of highly oriented pyrolytic graphite (HOPG) with high quality as shown by the nanoscale morphology in Fig. 1b. Details of epitaxial growth and corresponding characterization are given in Methods and section A of Supplementary Information (SI). Fig. 1c shows an atomic scale STM image of YbCl$_3$ with typical atomic defects where the triangular array of protrusions represents the topmost Cl-lattice. The absence of a single protrusion leaving an individual hole in the image is identified as a Cl-defect (marked by the dashed red circle), and the triangle depression, upon which the trio of protrusions remains, is assigned to a vacancy of Yb-ion beneath the topmost Cl-layer (marked by the dashed black circle).

To investigate the electronic structure, tunneling spectra of YbCl$_3$ are obtained. The top panel in Fig. 1d shows the dI/dV spectrum (blue curve) taken at a relatively large tip-sample distance (the small distance case is described down below), unveiling the insulating nature by the full gap with vanished tunneling conductance around the Fermi level. Density functional calculations of sole YbCl$_3$ monolayer produce density of states (violet and orange curves displayed in Fig. 1d) that nicely agree with the dI/dV spectrum. The conduction band (the states detected from 0.8 to 2.0 eV) is the upper Hubbard band derived from the unoccupied Yb-4f orbitals while the valence band (states detected below -0.8 eV) stems from the occupied Cl-3p orbitals. The lower Hubbard bands of occupied Yb-4f orbitals subjected to a significant Coulomb correlation are buried deep below the Fermi level at −9 eV to −5 eV (see Fig. S1) which exceeds the accessible energy range of tunneling spectroscopy. As for now, no state is observed within the full gap range from -0.8 V to 0.8 V. Henceforth, we use these two explicit values to denote the boundary of the insulating gap and define the bias voltages from -0.8 V to 0.8 V as in-gap range (IGR), while other voltages beyond the full gap belong to out-of-gap range (OGR).

Surprisingly, unexpected non-zero differential conductance (coined as in-gap states later in the text) arises as several humps added to a parabola-like background within the IGR (bottom panel of Fig. 1d) when dipping the tip closer to the surface. Whereas, no density of states shall exist in IGR according to the DFT computation. The in-gap dI/dV curve is quite distinct from the typical V-shape spectrum of graphene, excluding the simple explanation of tunneling into the graphene states directly through the YbCl$_3$ insulating barrier. The shape of the in-gap differential conductance is highly dependent on the orbital-selective tunneling matrix determined by the setpoint and tip apex, suggesting the multiorbital nature of the in-gap states (see details in section E of SI).

Moreover, many peculiarities of the in-gap states are observed as shown in Fig.1e, f. Firstly, superlattices with different lattice constants arise and become particularly prominent as the bias approaches and enters IGR. The periodic corrugation resolved in IGR (right panel of Fig. 1e, f) is much stronger than that resolved in OGR (left panel of Fig. 1e, f). The differential conductance mapping (Fig. S2) illustrates the evolution of superlattice with bias and further proves that the superlattices are formed by the in-gap states. Secondly, the topmost Cl- and Yb- defects, though clearly resolved in OGR, become completely invisible as varying the bias voltage from OGR to IGR (Fig. 1e, f and Fig. S2), indicating the defect-immunity of in-gap states (the origin of the superlattices and the reason of their immunity towards ionic defects will be discussed later). Lastly, despite that all the in-gap states of different domains share the common features of superlattice and defect-immunity, their electronic properties are extremely domain-dependent. Domains with different superlattices, which show the same height in the image taken at OGR bias (left panel of Fig. 1f), exhibit a conspicuous height difference in the image taken at IGR bias (right panel of Fig. 1f, more data is given in Fig. S3) indicating the strongly different wavefunction distribution involved in the tunneling process. In summary, the pronounced phenomena of superlattice, defect-immunity, and domain-dependence in IGR indicate that the in-gap states are essentially distinct from the intrinsic states of $YbCl_3$ monolayer but originate from the $YbCl_3$/graphene interface. Similar in-gap states are also discovered in $MoS_2$/graphite moiré heterostructure that are attributed to the interfacial charge redistribution[36]. Later, we substantiate that the in-gap states are indeed interlayer charge transfer states and further characterize them as transferred electron-hole pairs which form Wigner crystals.

## Interlayer electron-hole pairs featuring Rydberg-like character

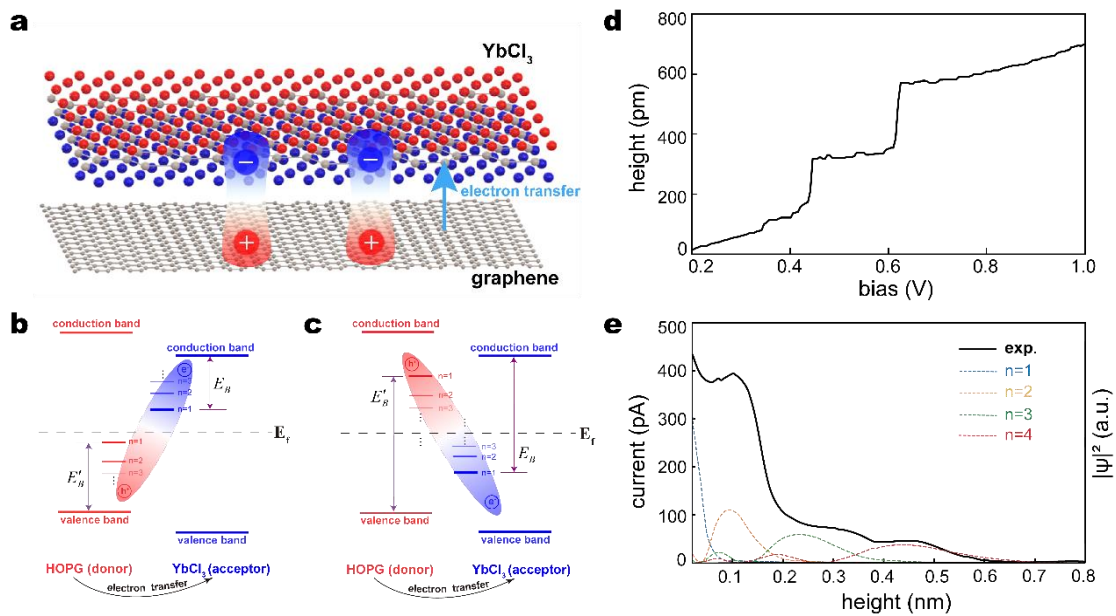

**Fig. 2. Charge transfer states at the interface. a**, pictorial depiction illustrating the electron transfer from the underlying graphene layer to the upper YbCl₃ layer. **b** and **c**, schematic plots describing the energy level configuration of the charge transfer states, where the Rydberg levels of transferred electrons and holes are represented by a column of blue and red bars respectively. The energy offset between charge transfer states and band edge scales with the binding energy (purple double arrows). **d**, the tip height-bias curve taken in constant-current mode at I=100 pA, showing a steplike feature which is a typical characteristic of the Rydberg-like series. **e**, the current-height characteristic curve measured at 0.7 V (black solid line). The colored dashed lines illustrate the spatial distribution of probability density for Rydberg-like states of $n = 1,2,3,4$.

Owing to the significant work function difference between two disparate materials, the YbCl₃/graphene heterostructure naturally leads to a charge transfer interface where electrons transfer from the underlying graphene layer to the YbCl₃ monolayer (see Methods), leaving planar electron- and hole- polarons residing in YbCl₃ monolayer and graphene layer respectively (Fig. 2a). The interlayer electron-hole pairs also conform to the definition of interlayer charge transfer excitons[37]. The transferred electrons and holes are bound to the interface by their mutual Coulomb attraction, forming Rydberg-like series[37,38]. Fig. 2b and 2c are the schematic diagrams of electron/hole levels in this charge transfer interface, the former illustrates a typical energy configuration of charge transfer interface[38] while the latter represents a case with large binding energy ($E_B$ and $E_B'$). Considering the vanished density of states at the Fermi level for the semimetal graphene, we simplify its electronic structure as a donor-type semiconductor. The electrons are transferred from graphene to the conduction band of YbCl₃ (the black arrow indicates the electron transfer) resulting in the same Fermi level across the interface. As shown in Fig. 2b, c, the $n = 1$ ($n$ is the principal quantum number of Rydberg-like series) energy position of transferred electron (hole) states is lower (higher) than the conduction band of YbCl₃ (valence band of graphene) by a binding energy $E_B$ ($E_B'$). If the binding energy is strong enough, it would make some hole (electron) states energetically higher (lower) than the Fermi level (as illustrated in Fig. 2c, which is confirmed to be our case, see section B and C of SI).

The Rydberg-like nature of the electron-hole pairs is unequivocally reflected in the height-bias (where the term "height" refers to the tip height here and hereafter) and current-height characteristic measurement (the experimental details are described in Methods). Firstly, as shown in Fig. 2d, by recording the tip height during ramping the bias voltage at constant-current mode, a steplike behavior appears in the height-bias curve, which is a well-known signature of Rydberg-like series in STM studies and is commonly observed in the Rydberg-like image potential states where electrons in front of a metal surface are trapped by the screening mirror charge[39-41]. Secondly, the evanescent shape of charge clouds along the direction normal to the interface towards the vacuum can be profiled by the current-height characteristic measurement. The current-height curve as shown in Fig. 2e exhibits an unusual oscillation with several nodes, pointing out that the current can even increase as retracting tip away from the surface. This observation is distinguished from a normal bulk band for which the current-height curve displays a monotonic decay (see section D of SI). We compare the current-height curve to the analytical spatial expression of the Rydberg eigenstates subjected to a one-dimensional Coulomb trap potential according to reference 39, assuming that the transferred electrons (holes) feel the attractive Coulomb potential provided by the planar transferred holes (electrons). It is found that the oscillating shape of the current-height curve can be explained by the superposition of probability clouds involving several Rydberg-like eigenstates of different $n$ (colored dashed lines in Fig. 2e). Nevertheless, we do not pursue a precise fitting of the experimental curve with the analytical expressions of Rydberg states because the coefficient of each $n$ item is determined by the orbital-selective tunneling matrix which is highly dependent to the setpoint and tip apex (see section B and E of SI). Moreover, the energy-resolved spatial distribution of charge transfer series is revealed by the bias-dependent current-height measurement. As altering the bias, the shape evolution of current-height curves helps to figure out the sequence of $n$ in energy space, confirming the large binding energy regime illustrated in Fig. 2c to be our case (see details about the measurement and tunneling mechanism in section B and C of SI). Therefore, it is the hole-like charge transfer states are probed at positive bias. As the fingerprints of Rydberg-like nature, the steplike shape in height-bias characteristic and the oscillating shape in current-height characteristic are prevalently present for the charge transfer states in IGR of all domains. While Fig. 2d, e shows the data taken at positive bias within IGR, the negative bias measurements probing the electron-like charge transfer states is shown in section B and F of SI, which exhibit a similar Rydberg-like nature.

**Wigner periodicities conforming to moiré patterns or commensurate with ionic lattice**

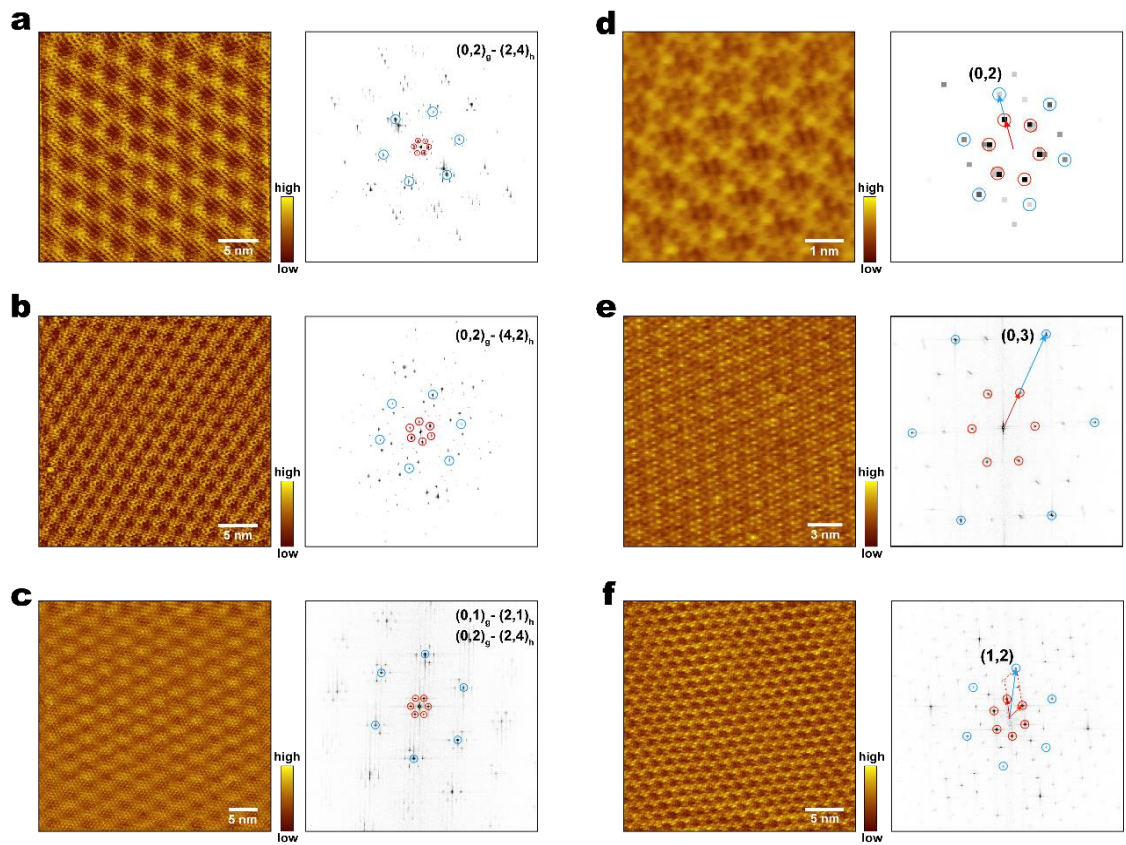

**Fig. 3. Wigner periodicities.** Real-space images and corresponding Fourier transformed maps of superlattices for charge transfer states. In the Fourier transformed maps, the reciprocal spots of $YbCl_3$ ionic lattice and superlattices are marked as blue and red circles, respectively. In **a-c**, the periodicities conform to moiré patterns. The legends point out the order of two spatial frequencies generating the moiré periodicity.

See details in section G of SI. In **d-f,** the superlattices are commensurate to the YbCl$_3$ ionic lattice with the precise fulfillment of the commensurability relations $\mathbf{h} = m\mathbf{K}_1 + n\mathbf{K}_2$, where $\mathbf{h}$ is the reciprocal wavevector of YbCl$_3$ ionic lattice, $\mathbf{K}_1$ and $\mathbf{K}_2$ are two basic reciprocal wavevectors of the superlattice, $m$ and $n$ are two integers so we henceforth use $(m, n)$ to denote the commensurability. Setup condition: U=0.6 V, I=500 pA for **a**; U=0.6 V, I=30 pA for **b**; U=0.7 V, I=100 pA for **c**; U=-0.8 V, I=-140 pA for **d**; U=0.6 V, I=200 pA for **e**; U=0.8 V, I=500 pA for **f**.

The results presented in the "Charge transfer states inside YbCl$_3$ band gap" section illustrate the prominent super-periodicities and defect-immunity exhibited by in-gap states. Since the in-gap states are unraveled to be the interfacial charge transfer states, the corresponding superlattices are thus formed by interlayer electron-hole pairs and evince the real-space arrangements of the latter. A rich variety of superlattices with well-defined superlattice constant are discovered experimentally as displayed in Fig. 3 and Fig. S4. It is natural to assign these superlattices to moiré patterns at first sight, which is verified to be true for some cases (see Fig. 3a-c and Fig. S4a-c, the analysis of moiré periodicity is described in section G of SI). However, a large number of superlattices are validated as not the case. As shown in Fig. 3d-f, the hexagonal superlattices exhibit a precise commensurate relationship with respect to the surface ionic lattice of YbCl$_3$ layer. The corresponding Fourier transformed maps straightforwardly demonstrate the exact fulfillment of the commensurability condition. Following the commensurability relation, the superlattices displayed in Fig.3d-f are named as (0, 2), (0, 3), (1, 2). There are more commensurate cases displayed in Fig. S4d-f, which cannot be generated by the moiré mechanism (see section G of SI). After an extensive search, all the discovered superlattices are either commensurate to YbCl$_3$ ionic lattice or in accord with a moiré periodicity. The classification is displayed in section G of SI. Although it is the hole-like transferred states which are imaged at positive bias, its spatial distribution synchronizes with the electron part. As demonstrated in Fig. S5, the periodicity of superlattice stays invariant under different bias voltages which also excludes the possibilities of quasiparticle interference pattern or charge density wave generated by Peierls mechanism. The superlattices with ionic lattice commensurability signify that the transferred electron-hole pairs are accordingly settled into the periodic dips of $YbCl_3$ ionic potential in a commensurate manner, which is a typical signature of Wigner lattice[18,19,42].

The distinguished phenomenon of defect-immunity which is present for all superlattices (except the special case of quasicrystalline phase discussed later) highlights the solid nature of Wigner crystal. The vanishment of Cl- and Yb-defects in these superlattices indicates the interaction-induced incompressibility of Wigner solid where the Coulomb-driven insulating electronic states are resilient and robust against ionic defects or other electrical perturbations[20]. The impingement caused by the ionic defects is to pin down the lattice sites of Wigner solid[43,44] instead of breaking its intactness. Lacking a Fermi surface consisting of extended Bloch electron states that can be scattered, all the electronic wavefunctions in the solid phase are exponentially localized in Wigner sites[11] which results in a significant insensitivity towards defects.

The defect-immunity nature demonstrates that all superlattices, i.e., both the moiré and the commensurate type, are incompressible Wigner solids. The existence of superlattices of two types indicates a mechanism where the moiré potential and ionic lattice potential compete for trapping the transferred electrons, and the final superlattice depends on whether the electrons are settled according to moiré potential or $YbCl_3$ crystal potential. In addition, as the defect-immunity is based on the incompressbility of Wigner solid, it will be lifted as soon as the electron solid begins to melt, which will be discussed in the next chapter.

**Wigner quasicrystal**

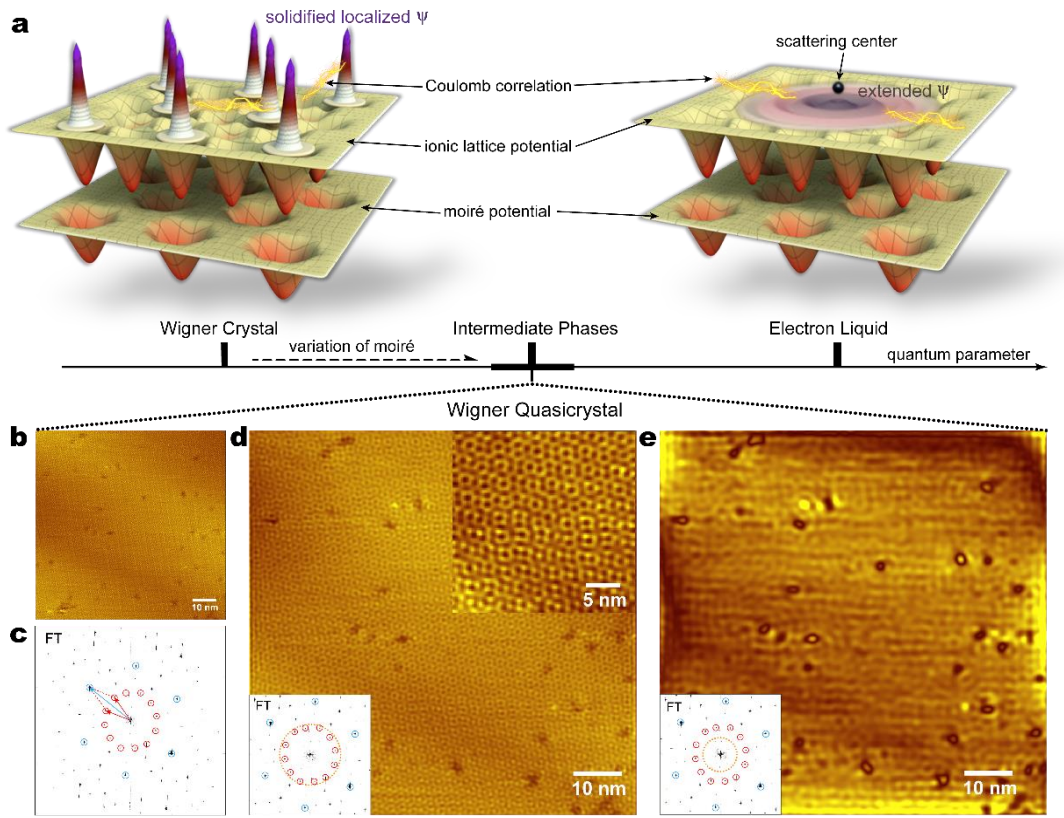

**Fig. 4. Quantum melting and Wigner quasicrystal. a,** schematic drawings of Wigner crystal and electron liquid in the moiré interface. The crystal and liquid phases are two ends of the quantum phase transition subjected to the quantum parameters of ionic lattice potential, moiré potential and Coulomb correlation. The electron wavefunctions are exponentially localized in corresponding Wigner sites in the imcompressible Wigner crystal, while the liquefied wavefunctions are extended and can be scattered in the electron liquid phase, resulting in the interference pattern. The variation of moiré parameter brings about the quantum melting of Wigner solid and the discovery of quasicrystalline melting phase. **b**, the STM image of quasicrystalline phase. Setup condition: U=0.6 V, I=300 pA. **c**, the Fourier transformed map of **b** where a dodecagonal motif of the superlattice is presented (red and blue circles indicate the reciprocal spots of supermodulation and $YbCl_3$ lattice, respectively). The commensurability relation is also precisely fulfilled here as marked by the red and blue arrows. **d**, the texture of Wigner quasicrystal obtained by filtering out spatial frequencies higher than the dodecagonal superlattice. The top-right inset displays the close-up of quasicrystalline texture. In the left-bottom inset, the dashed orange circle

indicates the low-pass filtering mask. **e**, Fourier-filtered image illustrating the quantum interference pattern. In the inset, the dashed orange circle indicates the low-pass filtering mask.

The collaboration and competition between ionic lattice, moiré periodicity, and density of correlated charges might cultivate a Coulomb frustration regime where the quantum phase transition such as the quantum melting of Wigner crystal could occur. The all-encompassing orientations of epitaxial monolayer domains develop a wealth of moiré landscapes, providing the degree of freedom of the moiré quantum parameter to explore the quantum phase transition (Fig. 4a). By investigating among different domains, the melted Wigner orders of quasicrystalline phase is discovered.

As shown in Fig. 4b, the charge transfer states of certain domains, exhibit a mottled pattern rather than a well-defined lattice structure. The real-space image is disordered, however, a remarkable delicate motif is unveiled after Fourier transformation. As shown in Fig. 4c, the Fourier transformed map exhibits a regular dodecagon of reciprocal spots of the superlattice (twelve points marked by the red circles). Similar to the commensurate Wigner lattices described above, the addition of two neighboring reciprocal vectors of supermodulation $K_1, K_2$ is equal to the reciprocal vector of $YbCl_3$ ionic lattice $h$ (marked by the red and blue arrows in Fig. 4c), which signifies the Wigner-crystalline commensurate arrangement of electrons. The beating frequencies of $K_{1,2}$ and $h$ produce another six regular dodecagons centering at the reciprocal spots of $YbCl_3$ ionic lattice (six blue circles). As an expression of the commensurability, the seven dodecagons organize in a corner-shared manner where every two dodecagons share the common two points. After filtering out the spatial frequencies larger than the dodecagonal peaks (the filtering window is displayed in the inset of Fig. 4d), the real-space image of supermodulation reveals a quasicrystal texture. Lacking a translational order but retaining the 12-fold orientational order, the real-space texture and Fourier transformed motif is exactly identical to the atomic dodecagonal quasicrystal[45], proving that, besides the quasicrystal made of atoms, the melting of Wigner crystal can also lead to Wigner quasicrystals made of electrons.

Moreover, the defects, which are invisible in other Wigner lattices of the solid phase, reappear in the quasicrystalline lattice here. This indicates the onset of melted electron liquid with extended Bloch states that can be scattered by defects, resulting in an interference pattern illustrated in the filtered image (Fig. 4e). As a result, the dodecagonal quasicrystalline supermodulation observed here is proved to be an electron liquid-solid hybrid where the quantum nature plays a more important role compared to atomic quasicrystal. The solid characteristic manifests in the commensurability displayed in Fourier transformed map, while the liquid characteristic is unveiled by the interference pattern caused by scattering. As scattering centers, most of the ionic defects resolved in charge transfer states here do not coincide with the positions of topmost Cl- and Yb-defects imaged at OGR, and show a greatly reduced density compared to the latter (see Fig. S6), suggesting that they are bottommost Cl-defects. This observation is in line with the expectation since the interfacial confinement of charge transfer states makes them more sensitive to the defects close to the interface.

## Discussion and outlook

The amount of charge transfer is around 0.21 e/nm$^2$ from the computational estimation (see Methods). In other words, the average interelectron distance is a few nanometers, exactly the same order of magnitude compared to the observed superlattice constant, making the various competing parameters comparable and leading to a rich phase diagram. The critical melting distance between two neighboring electrons in an ideally clean two-dimensional system is predicted to be $r_s$=37, in the unit of $a_0$ ($a_0$ is the Bohr radius so that $r_s$ is roughly 2 nm)[2]. However, $r_s$ varies from 5 to 60 in realistic two-dimensional electrons[7] which is close to the interelectron distance in our system. We suggest that the accordant magnitudes make the quantum melting and ensuing Wigner quasicrystal accessible in our system.

The Wigner solid observed here is constructed on the two-dimensional bound electron-hole pairs in YbCl$_3$/graphene interface. However, without seeing Wigner lattices commensurate with graphene lattice, only the ones commensurate with the YbCl$_3$ lattice are present, indicating the dominance of transferred electrons. The arrangement of holes in graphene, owing to the drag from Coulomb attraction, synchronizes with the electrons as reflected by the same superlattice in positive and negative bias. Due to the significant localization of 4f orbitals, rare-earth elements often induce numerous correlated phenomena, such as RKKY interaction, Kondo physics, and non-Fermi liquid[46]. Therefore, this is a reasonable observation in light of the tremendous Coulomb correlation of transferred 4f electrons. In addition, the charges in graphene layer cannot form Wigner crystal, as the kinetic energy of filled charges scales linearly with the electrostatic energy due to the specific Dirac dispersion[4].

We expect that more insight could be gained by further introducing the moiré Hamiltonian into the Hubbard model in theory. However, a crucial issue arises that the basic unit here is the electron-hole pair where the electron-part dominates. Depending on the comparison between interlayer electron-hole spacing and lateral interparticle spacing, a fermi-boson crossover would occur which decides either a Bose-Hubbard model or Fermi-Hubbard model shall one adopt, leaving the possibility of BCS-BEC crossover[30]. More exotic quantum states are prospective to be achieved and uncovered here via combined experimental techniques. In summary, the f-orbital halides/semimetal charge transfer interface provides a fertile playground of many-body physics, demanding further experimental and theoretical exploration.

## Methods

**MBE growth**

The YbCl$_3$ monolayer was grown on cleaved HOPG substrate by MBE at a base pressure of ~ $1\times10^{-9}$ mbar. The commercial anhydrous YbCl$_3$ powder was evaporated at 680 °C and the substrate was kept at 340 °C during the growth. The process of growth was monitored by reflective high-energy electron diffraction (RHEED). The YbCl$_3$ monolayer coverage reaches roughly 60% after 30 minutes of deposition (see section A of SI). In this work, we focus on a sample with full coverage of YbCl$_3$ monolayer which is realized after 50 minutes of deposition.

**STM measurements**

Experiments were performed with a home-built hybrid STM and MBE system, and the STM measurements were operated at a temperature of 5 K. The sample was directly transferred into STM to perform in situ measurement after finishing the growth. Scanning tunneling spectroscopy was performed using a lock-in amplifier technique with a modulation frequency of 963 Hz and a root-mean-square modulation voltage between 5 mV and 20 mV depending on the spectral range of interest.

**DFT calculation**

The density functional theory calculations for YbCl$_3$ monolayer are carried out using the full-potential augmented plane wave plus local orbital code (Wien2k)[47]. We use the experimental lattice parameters of a=b=6.64 Å and optimize the atomic positions. The plane wave cutoff energy of 14 Ry is set for the interstitial wave functions, and 11*11*1 k-mesh is used for integration over the Brillouin zone. The muffin-tin sphere radii are chosen to be 2.5 and 2.0 Bohr for Yb and Cl atoms, respectively. We perform LSDA+U+SOC calculations with the typical values Hubbard U=8.5 eV and Hund exchange $J_H$=1.0 eV for Yb 4f electrons[48], and the SOC effect is included by the second-variational method with scalar relativistic wave functions.

**Height-bias characteristic**

We perform the height-bias characteristic measurement by recording the tip position during ramping the bias voltage in constant-current mode, where a feedback system raises the tip height to maintain the current as more tunneling channels are opened by the increased bias. The start point of the tip position determined by the initial setup condition is considered as the zero-height point. The salient steplike feature is commonly recognized as the fingerprint of Rydberg-like series seeing that the tip-sample distance is exponentially sensitive to the energy threshold of quantum states.

**Current-height characteristic**

Before we take the current-height characteristic, the setup condition described in Fig. 2e is adopted so that the tip is positioned over the sample at a relatively close distance in order to ensure that the interfacial charge transfer states can be accessed by the tip. Then disable the feedback to fix the tip at current position which acts as the zero-height start point in the curve recorded later. Next step is to retract the tip and, in the meantime, record the tunneling current so the current-height curve is obtained. The bias can be set to the chosen value in the backward process of the tip, therefore the bias-dependent current-height characteristic can be acquired.

**Work function and interfacial charge transfer**

The work function of $YbCl_3$ monolayer is calculated to be 8.2 eV, which is larger than that of 4.6 eV for graphene, suggesting a charge transfer across the $YbCl_3$/graphene interface. In our computation, $YbCl_3$ monolayer is slightly electron doped when contacting to graphene, and the electron doping level is estimated to be 0.21 $e/nm^2$ by Bader charge analysis[49]. In realistic system, the electrostatic field of the various two-dimensional Wigner crystals formed in the interface would alter the work function difference and the consequential amount of charge transfer accordingly.

**Data availability**

The data that support the findings of this study are available from the corresponding authors upon reasonable request. Source data are provided with this paper.

**Acknowledgements** We thank Wulf Wulfhekel and Xiaopeng Li for insightful discussions. C. G. acknowledges funding from the National Key Research and Development Program of China (Grant No. 2019YFA0308404), Science and Technology Commission of Shanghai Municipality (Grant No. 20JC1415900) and Shanghai Municipal Science and Technology Major Project (Grant No. 2019SHZDZX01). H. W. acknowledges support from the National Natural Science Foundation of China (Grants No. 12174062). F. Y. is sponsored by Shanghai Pujiang Program No. 19PJ1401000 and National Natural Science Foundation of China (Grant No. 12004076).

**Author contributions** M.Z. fabricated the samples. M.Z., Z.W. and C.W. performed the measurements. L.L. conducted the DFT computation. Z.W., M.Z., C.W., F.Y. and C.G. analysed the data. L.L. and H.W. conducted the computational analysis. Z.W. and C.G. conceived the project and wrote the paper with inputs from all co-authors.

**Competing interests** The authors declare no competing interests.

Supplementary Information for:

# Imaging the quantum melting of Wigner crystal with quasicrystalline order

Zhongjie Wang[1,2], Meng Zhao[1,2], Lu Liu[1,2,3], Chunzheng Wang[1,2], Fang Yang[4,5], Hua Wu[1,2,3,6]* & Chunlei Gao[1,2,4,5,6,7]#

*1 State Key Laboratory of Surface Physics and Department of Physics, Fudan University, Shanghai 200438, China*

*2 Shanghai Qi Zhi Institute, Shanghai 200232, China*

*3 Laboratory for Computational Physical Sciences (MOE), Fudan University, Shanghai 200438, China*

*4 Institute for Nanoelectronic Devices and Quantum Computing, Fudan University, Songhu Rd. 2005, Shanghai 200438, China*

*5 Zhangjiang Fudan International Innovation Center, Fudan University, Shanghai 201210, China*

*6 Collaborative Innovation Center of Advanced Microstructures, Nanjing University, Nanjing 210093, China*

*7 Shanghai Research Center for Quantum Sciences, Shanghai 201315, China*

*These authors contributed equally: Zhongjie Wang, Meng Zhao, Lu Liu*
*Corresponding authors: \*wuh@fudan.edu.cn, #clgao@fudan.edu.cn*

# Supplementary Text

## A. Characterization of surface morphology at different growth stage.

The coverage of YbCl$_3$ monolayer reaches roughly 60% after 30 mins deposition at the growth condition described in "Methods" section. The corresponding surface structure is characterized by RHEED and STM. As shown in the STM image of Fig. S7a, an YbCl$_3$ monolayer island of high quality with a height of 4 Å is formed on the HOPG substrate. The lattice constant of epitaxial YbCl$_3$ single layer is observed to be 6.64 Å, which is calibrated via the direct comparison between the lattice structure of YbCl$_3$ monolayer and HOPG substrate. The main text mentions that we focus on a sample of full coverage realized by 50 mins deposition. Fig. S7b shows the surface characterization of a sample of 70 mins deposition, which exhibits the same fully covered morphology of YbCl$_3$ monolayer, suggesting that prolonging the deposition duration at the same growth condition cannot lead to the formation of a second layer.

## B. Energy levels configuration at the charge transfer interface.

In the main text, Fig. 2b, c illustrates two configurations with binding energy of small size and large size, respectively. The electron (hole) series, whose principal quantum number ascends (descends) as increasing energy, lie above (below) the Fermi level in Fig. 2b. In Fig. 2c, the large binding energy leads to a different configuration where some hole (electron) series are pushed to energy positions above (below) the Fermi level. To find out the exact energy configuration of charge transfer states, we perform the bias-dependent current-height characteristics measurement. Controlling the tip to repeat the same path of retraction at different tip-sample voltage, the bias-dependent current-height curves are obtained. The results are shown in Fig. S8. The tunneling current is contributed by the integral of quantum states lying in between the Fermi level and the value of bias, which helps to decipher the tunneling components inside the energy range of integration. As shown in Fig. S8b, the current-height curve recorded at 0.8 V shows a composite shape containing both the monotonic decaying of small n (for example, n=1) state and the oscillating behavior of large n states. As lowering the bias close to the Fermi level till 0.1 V, the trend of monotonic decay, which corresponds to the portion with smaller n, is gradually reduced and replaced by the trend of a drastic oscillation with two nodes, indicating an exclusive portion of n=3 or n>3 states. Thus, smaller n states occupy higher energy positions and larger n states occupy lower energy positions in the positive bias range. This result demonstrates that a hole-sequence of Rydberg-like series lie above the Fermi level, of which the n descends with increasing energy. Vice versa, the ascending n of electron-sequence is also observed below the Fermi level (Fig. S8c). Thus, the configuration with large binding energy described in Fig. 2c (Fig. S8a) is confirmed to be our case. The large binding energy observed here can be ascribed to two reasons. Firstly, the remarkable localization of f-orbital polarons efficiently enhances the Coulomb correlation. Secondly, the metallic nature of underlying graphene renders a much tighter binding due to the high mobility of graphene hole states.

## C. Tunneling scheme for charge transfer states

We report the interfacial charge transfer states which appear as the in-gap states in tunneling spectrum featuring the Rydberg-like spatial distribution. Since charge transfer is a rather common phenomenon which occurs for almost all heterostructures, one would expect that such interfacial in-gap states are supposed to be universal in STM measurement. However, the relevant report is very seldom. From the standpoint of the tunneling mechanism, we think that the exact energy configuration of charge transfer states, i.e., the relative position of electron/hole levels with respect to the Fermi level, plays a crucial role in determining whether the charge transfer ground state can be detected by the tunneling probe or not. In Fig. S8, the energy configuration of Fig. 2c is experimentally pinned down where some electron series lie below the Fermi level and some hole series lie above the Fermi level. We think that this energy configuration is not merely an observation, but also a necessary requirement for producing the tunneling conductance by charge transfer states.

**I. Energy configuration of charge transfer states inaccessible to tunneling probe.** The small binding energy case as depicted in Fig. 2b is a more common one for interfacial charge transfer states, where the transferred electron states lie above the Fermi level with the $n=1$ state occupied by an electron in the ground state (similarly, the hole states occupy energy positions below the Fermi level). After taking the tunneling mechanism into account, however, the charge transfer states in this case are invisible to tunneling probe as they do not participate directly in the tunneling process. When setting the bias voltage to a positive value, the tip Fermi level is higher than that of the sample, and the electrons tunnel from the tip into the unoccupied states above the Fermi level of the sample where the transferred electron series are situated in Fig. 2b. While the electron-like series of different $n$ are excitation levels of the transferred electron which belongs to the electron-hole pair, they are not empty states to accept the electrons tunneling in, and thus they will not directly show up in the tunneling spectrum. Analogously, the hole series blow the Fermi level also do not contribute to the tunneling current, making the charge transfer states in this case invisible to tunneling probe. Nevertheless, the charge transfer states shall make a difference in the tunneling process, which manifests as a side effect. In terms of the dipole field derived from interlayer electron-hole pairs, the intermediate state after the injection of tunneling electron would be renormalized and turn out to be a three-particle state consisting of two electrons and a single hole, that is, the trion state[1]. In other words, the tip electron tunnels through an intermediate state where the tunneling electron is trapped by the dipole field of interlayer electron-hole pairs. Since the trion state formed after the electron injection is not the ground state, the tunneling electron would leave the trion state soon for the unoccupied graphene state, contributing to the tunneling current and leading to a tunneling peak in the conductance spectrum. Compared to the exciton, the typical binding energy of trion state is greatly reduced. Consequentially, such conductance peak induced by the trion state is probably located at the energy position tens of millivolts[2] below the conduction band minimum, with the charge-transferred Rydberg series invisible to the tunneling process.

**II. Energy configuration of charge transfer states accessible to tunneling probe.** As for the large binding energy case illustrated in Fig. 2c, things are different as the unique energy configuration renders the charge transfer states accessible via tunneling process. Since the hole series are situated above the Fermi level, they can accept the tunneling electrons at posive bias voltage, and vice versa for the electron series below the Fermi level. At positive bias, the tunneling electron annihilates the hole state and destroys an interlayer electron-hole pair. Since that the interlayer electron-hole pairs are energetically favored ground state, the tunneling electron which annihilates the electron-hole pair will leave the hole state rapidly owing to the fast interlayer charge transfer[3], restoring the interlayer electron-hole pair and giving rise to tunneling conductance. As long as the time of the fleeing of excess tunneling charge and the restoring of electron-hole pairs is much shorter compared to the tunneling lifetime, the effect of tunneling charge is pertubative and the pair annihilation is insiginificant. As a result, the charge transfer states in this case directly participate in the tunneling process and show up in the tunneling spectrum. In the measurement shown in Fig. S8, the electron tunnels out from the occupied charge transferred states of the sample at negative bias, displaying an ascending n sequence of electron series. When tunneling into the empty states of the sample at positive bias, the n sequence switches to the descending hole series immediately. Such sudden change with respect to the bias polarity is consistent with the tunneling regime where switching the polarity reverses the direction of electron tunneling at once. This result indicates that all the electron-like charge transfer series lying below the Fermi level are occupied (so do the hole-like states above the Fermi level), ensuring that there are electrons that could tunnel out from the occupied electron-like charge transfer states at negative bias, and could tunnel into the empty holes at positive bias. A corollary of Fig. S8 is that at least several charge transfer states of different n are occupied by electrons/holes, resulting in the bound state of multiple electrons and holes. We are not sure if each n state is doubly occupied or not, but the strong Coulomb repulsion is likely to favor a single occupation. The bound electron-hole states must contain multiple particles no matter what the internal electronic configuration is, generally speaking, making the multiparticle bound states more or less resemble Helium orbitals or Lithium orbitals. We attribute the unusual multiparticle bound states to the ordering and gathering process of transferred charges during the formation of Wigner lattice. Thus, very likely, more than one electron is localized at each Wigner site, which is consistent with the Wigner bubble phase[4].

**D. Current-height characteristic for intrinsic band of $YbCl_3$ monolayer.**

The unusual oscillation observed in the current-height characteristic is a representative feature of charge transfer states, proving the Rydberg-like nature. When increasing the tip-sample distance to probe the intrinsic band states of $YbCl_3$ monolayer, as expected, a common monotonic decay is observed in the current-height characteristic. Fig. S9 shows the results, the tip-sample distance is gradually increased by changing the setpoint bias from IGR to OGR while maintaining the setpoint current in constant-current mode.

### E. Multiorbital nature of tunneling process

When probing the charge transfer states, the tunneling process occurs between the tip states and the Rydberg-like series, while the latter contain many states of different n as well as of different spatial distribution normal to the interface. As revealed by the current-height characteristic in Fig. 2, the unusual shape of the current-height curve can be decomposed into several Rydberg-like components. Each component dominates different spatial area along the out-of-interface direction. Changing the tip-interface distance is supposed to alter the relative ratio of Rydberg components which participate in the tunneling process, leading to a distinct dI/dV spectrum. The result is shown in Fig. S10, where the current-height curve taken at 0.6 V in Fig. S10a shows a typical multiorbital nonmonotonic oscillation, featuring a hump around 0.2 to 0.3 nm and an undulation at the tail around 0.5 to 0.6 nm. Fixing the tip at different positions (marked as colored arrows) to perform dI/dV spectroscopy will lead to disparate tunneling spectra. Fig. S10b shows the corresponding dI/dV curves, which indicate the alternation of probed Rydberg components and explain the extreme setpoint sensitivity of dI/dV spectrum for charge transfer states. Similarly, changes in the tip geometry and anisotropy are also going to alter the orbital-selective tunneling matrix, resulting in a disparate shape of dI/dV curve.

### F. Steplike feature observed in height-bias characteristic taken at negative bias.

The Rydberg-like charge transfer states occupy the IGR in energy, rendering a steplike behavior in the height-bias characteristic across this energy range. While Fig. 2d illustrates the height-bias curve taken at positive bias, here, we show that the steplike feature is also present in the height-bias curve taken at negative bias (Fig. S11).

### G. Moiré periodicity

The possible moiré patterns formed in the heterostructure can be geometrically constructed by two beating spatial frequencies belonging to $YbCl_3$ lattice and underlying graphene lattice, respectively. Based on the comprehensive mathematical model[5] which involves not only the first order but also higher-order spatial frequencies of two coinciding lattices, the analytical expression of possible moiré wavelength and the relative angle with respect to the $YbCl_3$ lattice are obtained. We use $\mathbf{g}_1$ and $\mathbf{g}_2$ to denote the two reciprocal vectors of graphene lattice and $\mathbf{h}_1$ and $\mathbf{h}_2$ to denote the two reciprocal vectors of $YbCl_3$ lattice, so

$$|\mathbf{g}_{1,2}| = \frac{2\pi}{a} \tag{1}$$

$$|\mathbf{h}_{1,2}| = \frac{2\pi}{b}, \tag{2}$$

where $a = 2.46$ Å and $b = 6.64$ Å are the graphene lattice constant and $YbCl_3$ lattice constant respectively. Thus, the higher-order spatial lattice frequencies, which correspond to the higher-order spots in the reciprocal lattice, can be represented as the composition of the multiples of two reciprocal vectors:

$$\mathbf{g}_{(m,n)} = m\mathbf{g}_1 + n\mathbf{g}_2 \tag{3}$$

$$\mathbf{h}_{(r,s)} = r\mathbf{h}_1 + s\mathbf{h}_2, \tag{4}$$

where (m, n) and (r, s) tuples consisting of integers indicate the order of spatial frequencies of graphene and YbCl$_3$ lattice. Letting $\mathbf{K}_M$ be the reciprocal vector generated by the difference of $\mathbf{g}_{(m,n)}$ and $\mathbf{h}_{(r,s)}$, so we have that

$$\mathbf{K}_M = \mathbf{g}_{(m,n)} - \mathbf{h}_{(r,s)}. \tag{5}$$

Thus, varying the order tuple of (m, n) and (r, s) leads to the calculated $\mathbf{K}_M$ of different kinds, which gives the wavelength and orientation of the final moiré periodicity. However, there is no need to go through all the combinations of (m, n) and (r, s) because, on the one hand, the experimentally discovered superstructure has wavelength larger than the lattice constant of YbCl$_3$, which brings about the constraint upon the reciprocal vector

$$|\mathbf{K}_M| = |\mathbf{g}_{(m,n)} - \mathbf{h}_{(r,s)}| < |\mathbf{h}_{1,2}|. \tag{6}$$

This constraint condition greatly reduces the possible choices of (m, n) and (r, s). On the other hand, we shall focus on the low-order spatial frequencies with small integers in (m, n) and (r, s) since that the amplitude of higher-order spatial frequency is negligible. As a result, we find that the possible moiré periodicities generated by six low-order combinations listed down below contain all the non-commensurate superlattices discovered in experiment

$$(1,0)_\mathbf{g} - (2,1)_\mathbf{h}: \quad \mathbf{K}_M = \mathbf{g}_{(1,0)} - \mathbf{h}_{(2,1)} \tag{7}$$

$$(1,0)_\mathbf{g} - (1,2)_\mathbf{h}: \quad \mathbf{K}_M = \mathbf{g}_{(1,0)} - \mathbf{h}_{(1,2)} \tag{8}$$

$$(1,0)_\mathbf{g} - (3,0)_\mathbf{h}: \quad \mathbf{K}_M = \mathbf{g}_{(1,0)} - \mathbf{h}_{(3,0)} \tag{9}$$

$$(1,0)_\mathbf{g} - (0,3)_\mathbf{h}: \quad \mathbf{K}_M = \mathbf{g}_{(1,0)} - \mathbf{h}_{(0,3)} \tag{10}$$

$$(2,0)_\mathbf{g} - (4,2)_\mathbf{h}: \quad \mathbf{K}_M = \mathbf{g}_{(2,0)} - \mathbf{h}_{(4,2)} \tag{11}$$

$$(2,0)_\mathbf{g} - (2,4)_\mathbf{h}: \quad \mathbf{K}_M = \mathbf{g}_{(2,0)} - \mathbf{h}_{(2,4)}, \tag{12}$$

where $(m,n)_\mathbf{g} - (r,s)_\mathbf{h}$ points out the order of spatial frequencies participating in generating the moiré periodicity. Among the combinations of formula (7-12), cases (7-10) involve the lowest order spatial frequencies noting that $(1,0)_\mathbf{g}$ is the reciprocal vector of graphene lattice, while formula (11, 12) is the doubling frequency of formula (7, 8) respectively. Mapping the involved vectors into cartesian coordinate system and varying the relative orientation between graphene and YbCl$_3$ lattice allow us to establish the analytical functional relationship between the moiré wavelength and θ, where θ is the counterclockwise rotated angle modulo 60° between moiré lattice and YbCl$_3$ lattice. The function curve and datapoints of discovered superlattices are plotted in Fig. S12a, from which we can see that the superlattices marked as black boxes conform well to the moiré periodicities. In comparison, Fig. S12b plots the calculated datapoints of relevant commensurate superlattices with respect to YbCl$_3$ lattice together with the experimental datapoints. We can see that all the superlattices marked as red triangles, which are inconsistent with moiré periodicities (as shown in Fig. S12a), coincide with the commensurate-to-$YbCl_3$ superlattices. The two plots clearly illustrate that the observed superlattices can either be categorized into moiré type or commensurate-to-$YbCl_3$ type.

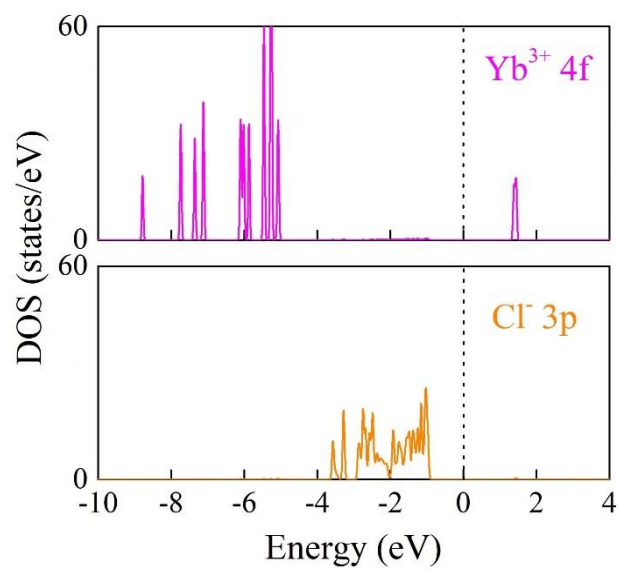

**Fig. S1 Computational partial density of states of sole monolayer YbCl₃.** The top panel illustrates the DOS of Yb-4f orbital, showing the upper Hubbard band (UHB) at 1.3 eV and the lower Hubbard bands (LHBs) from −9 eV to −5 eV. The bottom panel shows the DOS of Cl-3p orbital which constitutes the highest valence band.

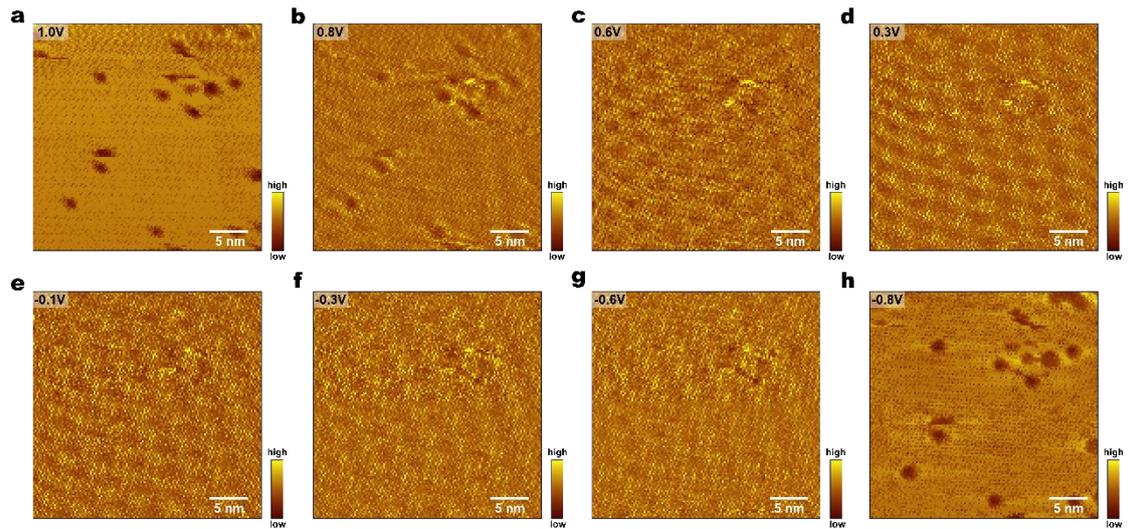

**Fig. S2 Energy evolution of in-gap states**. **a-h**, differential conductance maps taken at the same position in constant-height mode, from which we can see that the superlattice is prominent in IGR(c-g) and that the defects resolved within OGR (**a, b** and **h**) gradually vanish as entering IGR (**c-g**). The bias of each map is indicated in the upper left box.

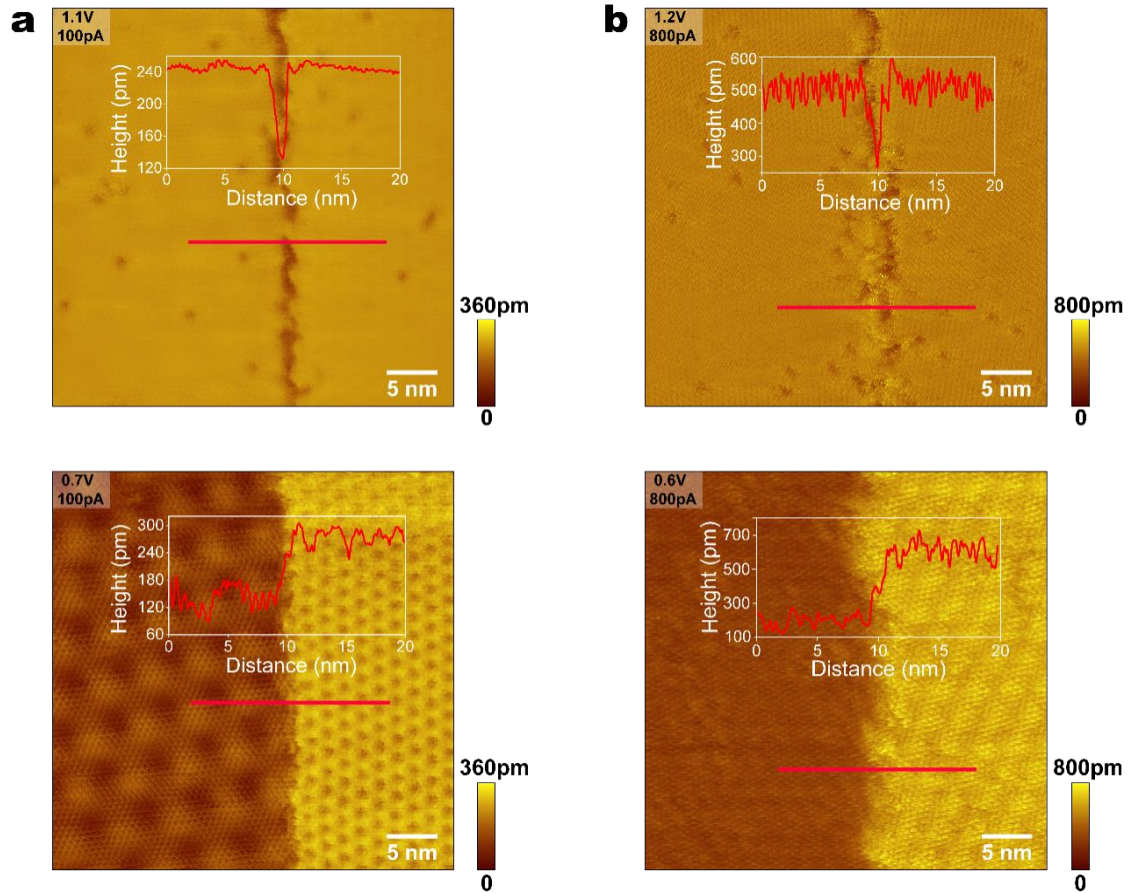

**Fig. S3 Intersection of monolayer domains imaged at OGR and IGR. a,** constant-current STM images taken at the same sample area displayed in Fig. 1f, but at a larger setup current I=100 pA (the setpoint of each map is indicated). The height difference between two domains observed in the bottom panel of **a** is about 1.5 Å, which is bigger than that of 0.8 Å in the right panel of Fig. 1f whose setup current is I=10 pA. The dependency of setpoint also suggests the multiorbital nature (see details in section E of SI). **b**, Images of another intersection of two domains with different superlattices. Similarly, no height difference is present in the image taken at OGR, while an ~3 Å height difference develops at IGR.

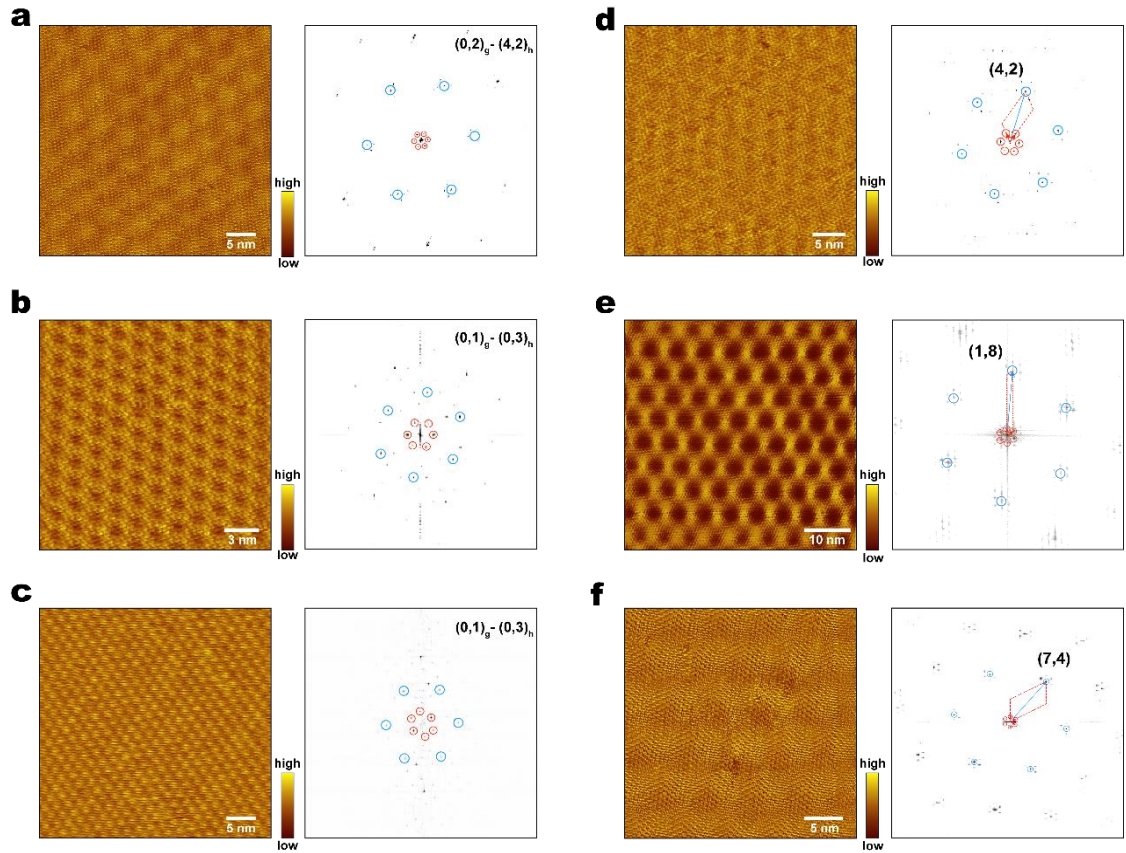

**Fig. S4 Other Wigner periodicities whose lattice structures are commensurate to the YbCl$_3$ ionic lattice or conform to the moiré periodicity.** **a-c**, real-space images and corresponding Fourier transformed maps of Wigner lattices whose periodicities are identical to moiré patterns. The legends point out the order of two spatial frequencies generating the moiré periodicity. **d-f** real-space images and corresponding Fourier transformed maps of Wigner lattices whose lattice structures are commensurate to the YbCl$_3$ ionic lattice. The commensurability relations are indicated. Setup condition: U=0.7V, I=200 pA for **a**; U=0.7 V, I=100 pA for **b**; U=0.6 V, I=2 nA for **c**; U=0.6 V, I=60 pA for **d**; U=0.7 V, I=100 pA for **e**; U=0.6 V, I=100 pA for **f**.

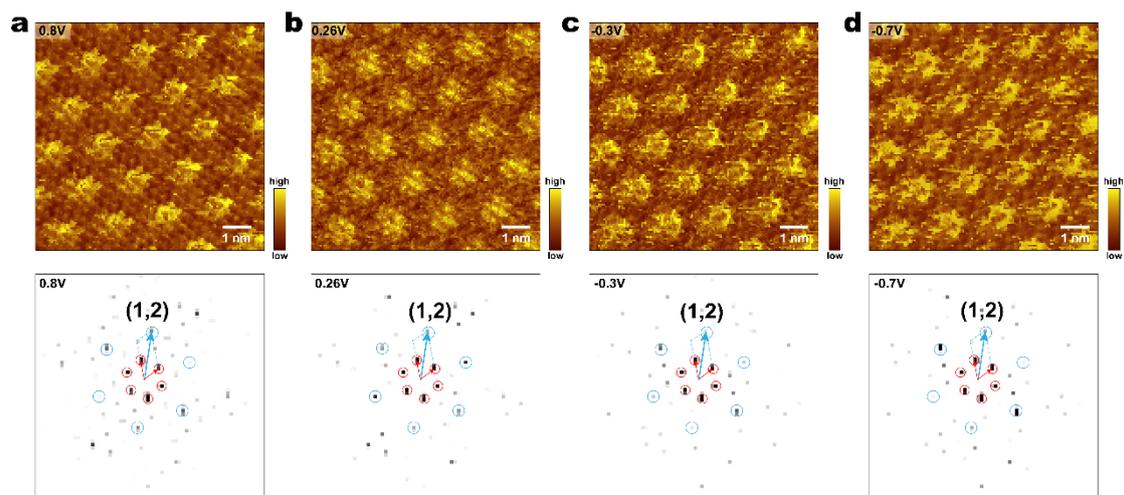

**Fig. S5 The superlattice imaged at different energy. a-d**, the bias-evolution of dI/dV map (top row) and its Fourier transformation (bottom row) of the (1,2) superlattice. The respective bias voltage is indicated at the top left of each image, and all the maps are taken at the same position, showing that the superlattice structure stays invariant with respect to different energy.

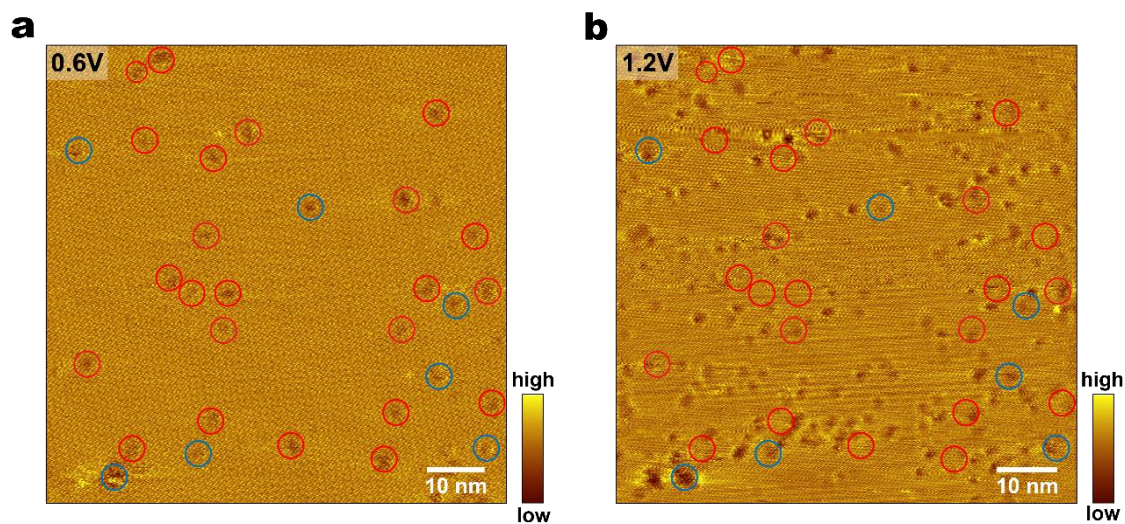

**Fig. S6 Defect positions resolved at IGR and OGR for Wigner quasicrystal. a**, STM image of quasicrystalline order taken at 0.6 V. **b**, STM image of the same position taken at 1.2 V. The blue and red circles mark the same positions of defects in **a** and **b**. Most defects imaged at IGR do not coincide with the defects resolved at OGR (which is the case for the defects marked by red circles), implying that they are bottommost Cl-defects.

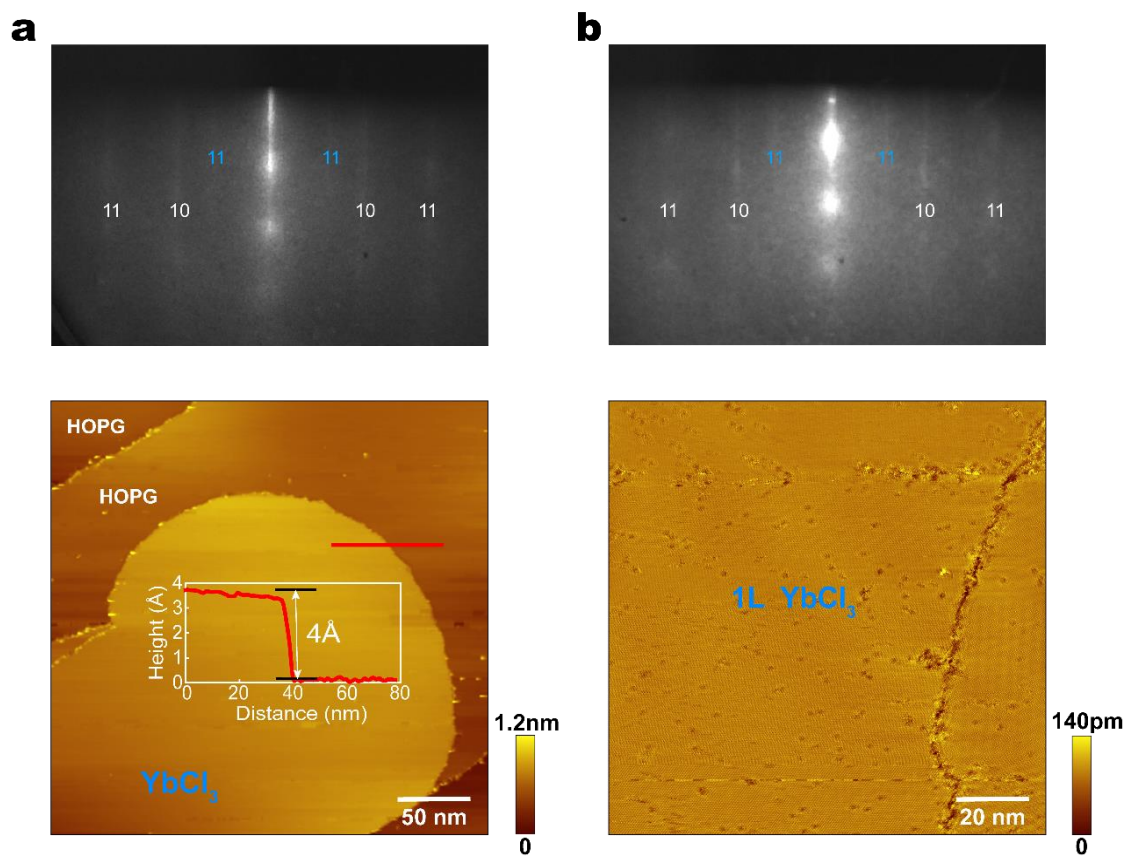

**Fig. S7 Surface characterization of samples with different growth durations**. **a**, RHEED pattern and STM image of the surface after 30 mins deposition. The white and blue numbers indicate the diffraction orders of RHEED patterns of HOPG and YbCl$_3$ monolayer, respectively. The height profile is presented in **a**. Setup condition for the STM image: U=3.4 V, I=30 pA. **b**, RHEED pattern and STM image of the surface after 70 mins deposition. Setup condition for the STM image: U=1.2 V, I=100 pA

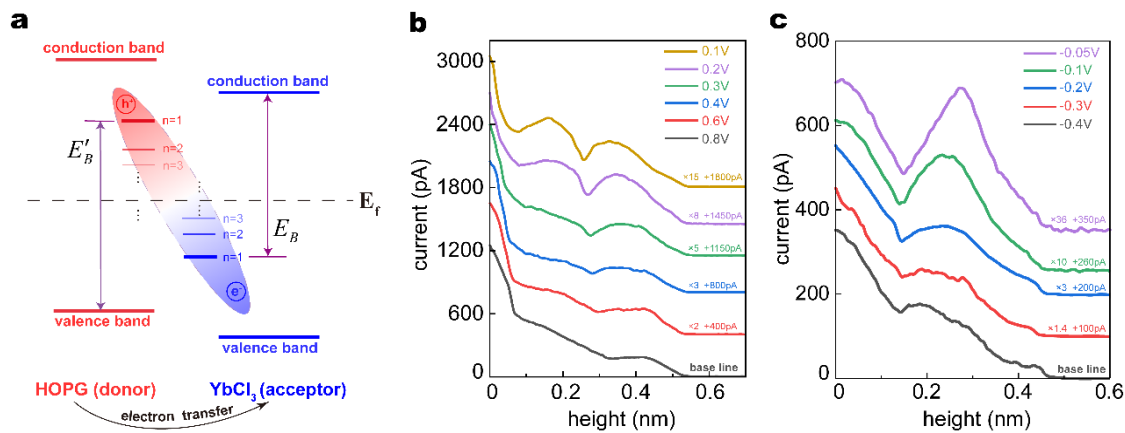

**Fig. S8 Energy-resolved normal spatial distribution of charge transfer states**. **a**, a plot same to Fig. 2c illustrating the energy level configuration with large binding energy. **b**, the current-height characteristic curves obtained at different positive bias voltages. All curves share the same zero-height point determined by the setup condition U=0.8 V, I=1.2 nA. For clarity, the lines from bottom to top are magnified and their zero-current points are upshifted with the factors indicated respectively. The shape evolution coincides with the hole-series whose n descends as increasing energy. **c**, the current-height characteristic curves obtained at different negative bias voltages, of which the shape evolution suggests electron-series whose n ascends as increasing energy. Similarly, the multiplicator and offset for each line are indicated.

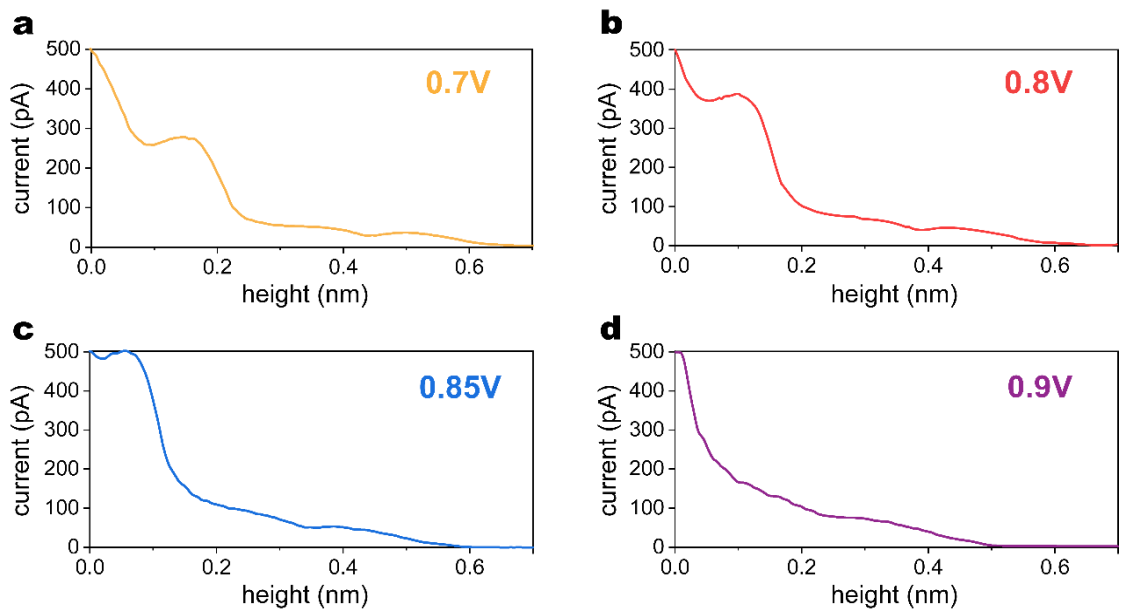

**Fig. S9 Evolution of current-height characteristic from charge transfer states to intrinsic band states**. **a-d**, current-height curves taken at different bias and different start point, the latter is determined by the corresponding setup condition: **a-d** share the same setup current I=500 pA, and the respective setup bias is indicated.

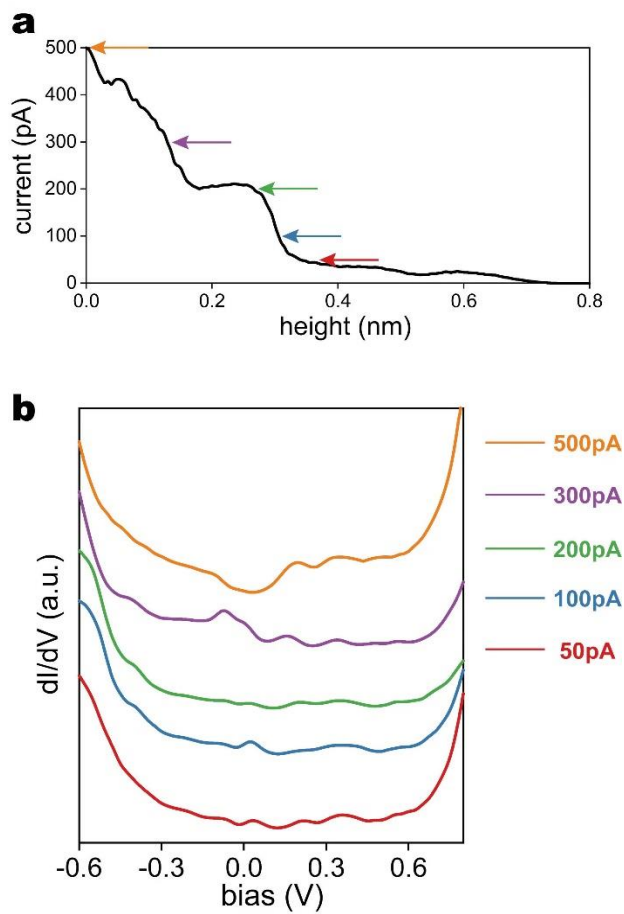

**Fig. S10 Tip-interface distance dependency of the tunneling spectroscopy for charge transfer states**. **a**, current-height curve taken at 0.6 V, the colored arrows indicate the positions at which the tip is fixed and the tunneling spectroscopy is performed. **b**, the corresponding dI/dV spectra taken at the positions marked in **a**. The legends indicate the setpoint current of each line.

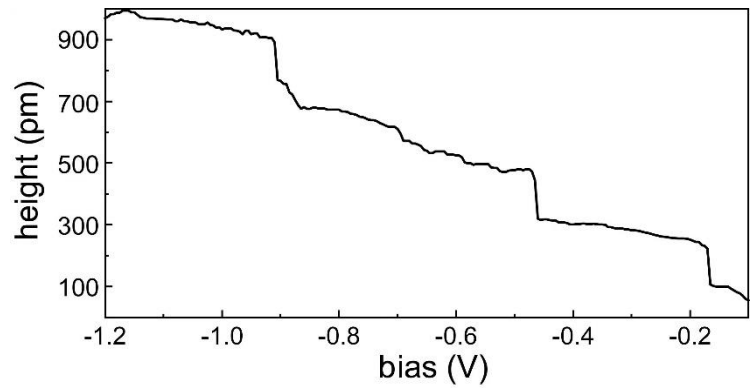

**Fig. S11 Height-bias curve taken in negative bias range with a fixed current of -350 pA.**

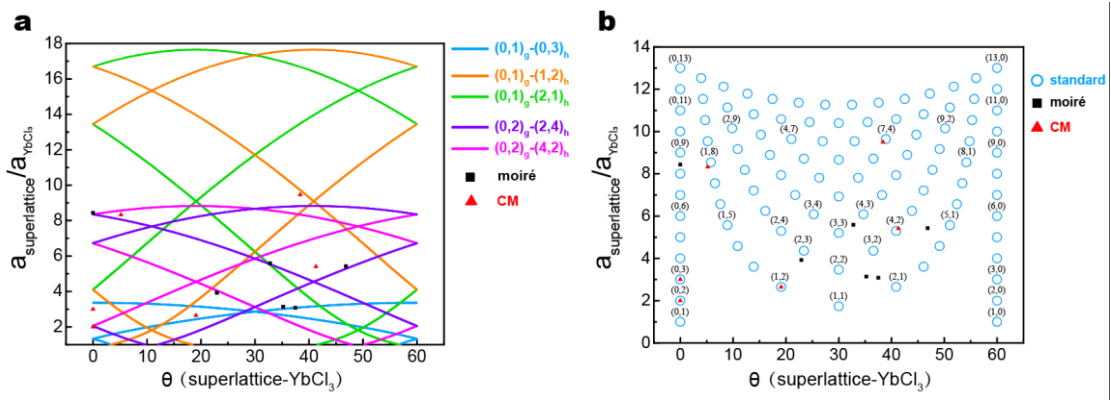

**Fig. S12 Classification of two types of Wigner lattices. a,** the analytical curve of possible moiré periodicities. The horizontal axis of θ is the relative angle between the moiré lattice and YbCl$_3$ lattice, the vertical axis represents the ratio of the superlattice constant to YbCl$_3$ lattice constant. The colored curves and legends indicate the order of spatial frequencies that generate the moiré periodicity. **b,** a plot illustrating all the standard θ and superlattice constant of the relevant commensurate-to-YbCl$_3$ superperiodicites. The blue circles represent the calculated coordinate points for standard commensurate (CM) periodicities. In comparison, all the data points for the experimentally observed Wigner lattices are indicated as solid boxes and triangles in **a** and **b**. The black boxes are the superlattices ascribed to moiré periodicities while the red triangles are those commensurate-to- YbCl$_3$ (CM) superlattices. Among the data points, the periodicity of (7,4) commensurate one is also very close to the $(0,1)_g$-$(1,2)_h$ moiré order (orange curve in the left panel), implying that the (7,4) Wigner lattice might originate from the collaboration of the ionic potential and moiré potential. The (0,2) commensurate one has small difference in periodicity with respect to neighboring moiré patterns (the pink and purple curves in the left panel), however, the high spatial resolution and precisely fulfilled commensurability relation (see in Fig. 3d) clarify its commensurate nature.